\documentclass[twocolumn,preprintnumbers,amsmath,amssymb,pre]{revtex4}

\usepackage{graphicx}
\usepackage{dcolumn}
\usepackage{bm}
\usepackage{color}
\usepackage{mathrsfs}

\def \rmd{\mathrm{d}}

\def \p{\partial}
\def \f#1#2{\frac{#1}{#2}}
\def \mr#1{\mathrm{#1}}
\def \l{\left}
\def \ri{\right}
\def \ds{\displaystyle}

\def \ds{\displaystyle}

\def \dd#1#2{\frac{\rmd#1}{\rmd#2}} 
\def \pp#1#2{\frac{\p #1}{\p #2}} 

\begin{document}
\title{Observation of particle acceleration in laboratory magnetosphere}

\author{Y. Kawazura, Z. Yoshida, M. Nishiura, H. Saitoh, Y. Yano, T. Nogami, N. Sato, M. Yamasaki, A. Kashyap and T. Mushiake }
\affiliation{Graduate School of Frontier Sciences, The University of Tokyo,
Kashiwa, Chiba 277-8561, Japan}

\date{\today}

\begin{abstract}
The self-organization of magnetospheric plasma is brought about by inward diffusion of magnetized particles.
Not only creating a density gradient toward the center of a dipole magnetic field, 
the inward diffusion also accelerates particles and provides a planetary radiation belt with high energy particles.
Here, we report the first experimental observation of a `laboratory radiation belt' created in the Ring Trap 1 (RT-1) device.
By spectroscopic measurement, we found an appreciable anisotropy in the ion temperature,
proving the betatron acceleration mechanism which heats particles in the perpendicular direction with respect to the magnetic field
when particles move inward.
The energy balance model including the heating mechanism explains the observed ion temperature profile.
\end{abstract}

\maketitle

\section{Introduction}\label{s:introduction}
Magnetospheres are natural plasma confinement devices ubiquitous in the universe.
Yet, their creation mechanism remains to be fully understood.
There must be a spontaneous mechanism that transports particles towards the center of a dipole magnetic field~\cite{Schulz,Dessler}.
However, such \emph{inward diffusion} seemingly violates the entropy principle, 
because it creates a gradient instead of flattening the density distribution.  
This challenge may be overcome by considering adiabatic invariants;
these are the magnetic moment, the bounce action and the longitudinal angular momentum.
Typically, the frequencies of the corresponding periodic motions are separated by three orders of magnitude.
The third invariant is most fragile, and its violation (i.e. the change of angular momentum) 
gives rise to the radial transport of particles.
When the diffusion of magnetized particles is constrained by the remaining adiabatic invariants,
the particle density tends to distribute uniformly on `magnetic coordinates' rather than on Euclidean coordinates.
Therefore, the homogeneous density on the former turns out to be inhomogeneous on the latter\,\cite{Birmingham,Hasegawa,YM2014}.

The same mechanism plays another interesting role, i.e., the acceleration of particles to produce a radiation belt~\cite{Kellogg,Brice,Carbary,Coroniti,Nishida}.
Conserving the remaining adiabatic invariants, 
the kinetic energy of a particle increases as it moves inwards.
An increase in cyclotron frequency (maintaining the magnetic moment constant)
results in betatron acceleration of the velocity in the direction perpendicular to the magnetic field, and an increase in bounce frequency (maintaining the bounce action constant) 
results in Fermi acceleration of the parallel velocity\,~\cite{Dessler}
(the electrons are further accelerated to an ultra-relativistic regime by whistler waves~\cite{Chen,Horne}).
Since the former is stronger than the latter, 
the inward diffusion results in temperature anisotropy~\cite{Dessler}.
Numerous satellite observations have provided evidence of anisotropic temperatures in planetary radiation belts. 
The anisotropic electron temperatures are consistent with the estimates from inward diffusion heating~\cite{whistler1,whistler2,whistler3};
ions also have anisotropic temperatures in the radiation belt of the Earth~\cite{Olsen1981,Olsen1987} as well as that of Saturn~\cite{Persoon}.

In addition to these theoretical and observational studies, the inward diffusion was experimentally verified recently.
On the Ring Trap 1 (RT-1) device which is a `laboratory magnetosphere' simulated by a levitated superconducting magnet~\cite{Yoshida2006,Yoshida2013}, the peaked electron density profiles were observed and proved to originate from inward diffusion~\cite{Saitoh2014}.
Similar density profiles were observed in the Levitated Dipole Experiment (LDX), which is another dipole confinement system~\cite{Boxer}.
On RT-1, distinct proof of inward diffusion was provided by producing a non-neutral (pure electron) plasma\,\cite{Yoshida2010};
by action-conserving acceleration, electrons diffuse into the central region where the electric potential is higher than their initial energies (such particles absorb energy from fluctuations, which drive radial motion).
Here, we report the first experimental observation of the other outcome of the inward diffusion, i.e. the particle acceleration.
Specifically, we observed the ion temperature anisotropy and identified its generation mechanism as betatron acceleration.
This finding means that we succeeded to demonstrate `laboratory radiation belt'.

\section{Experimental set-up}\label{s:set-up}
\begin{figure}[htpb]
	\begin{center}
		\includegraphics*[width=0.5\textwidth]{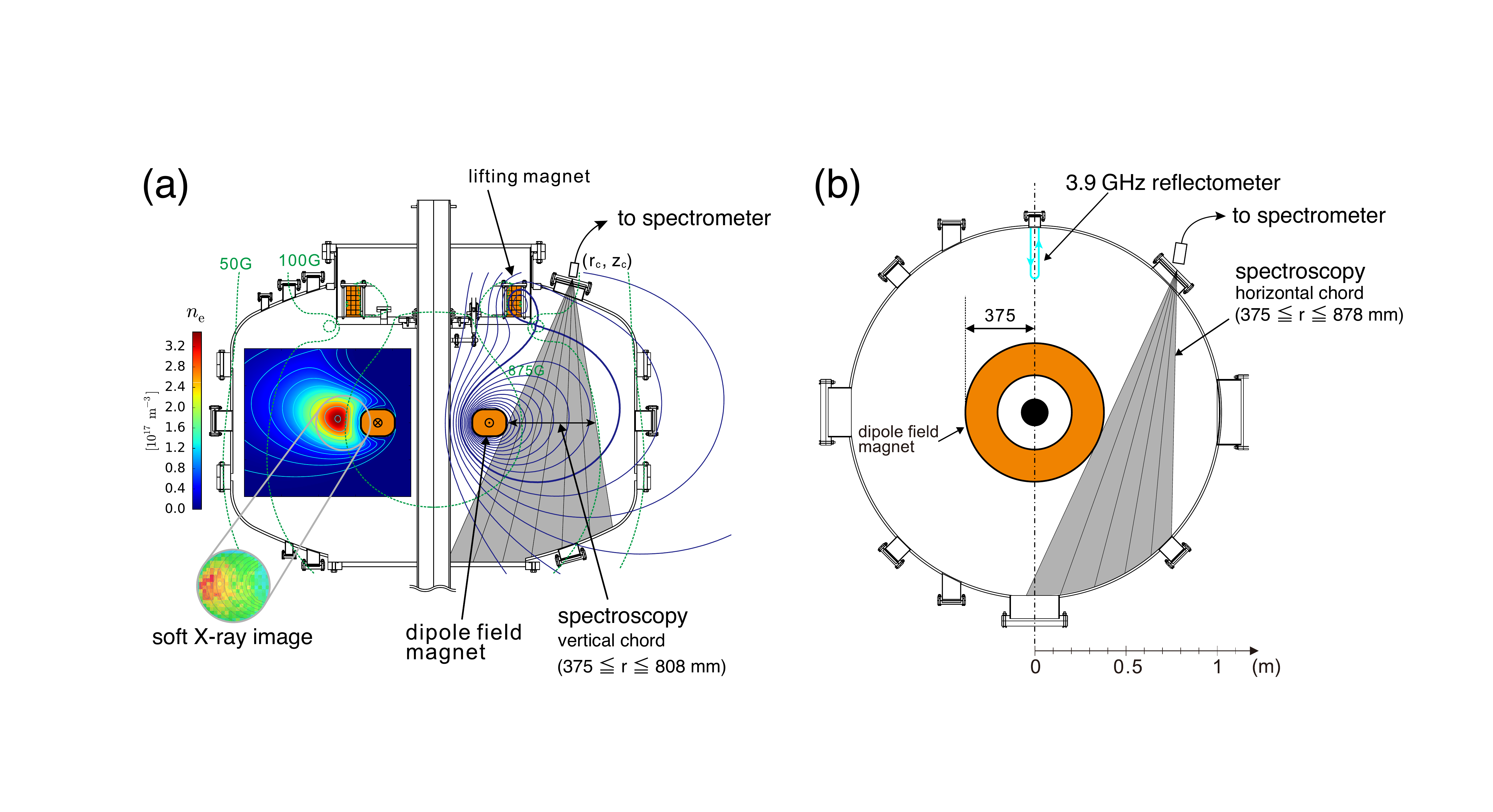}
	\end{center}
	\caption{ (a) The plan (vertical front view) of the device. 
						The superconducting magnet is levitated by the lifting magnet placed on the top of the vacuum chamber.
						The lines of sight of the vertical spectroscopic measurement cover the greyed vertical plane.
						On the left-hand side of the figure, the electron density profile
						 (measured by interferometers~\cite{Saitoh2014,Saitoh2015}) is plotted and the superimposed soft X-ray image (the interferometry measurement and the soft X-ray image are from different shots) is shown. 
						(b) The plan (top view) of the device. 
						The lines of sight of the horizontal spectroscopic measurement cover the greyed horizontal plane.
						The reflectometer (3.9\,GHz microwave) measures the density fluctuation
						near the cutoff density which is $\sim 1.9\times10^{17}\,\mr{m^{-3}}$ (light blue).  
					}
	\label{f:schematic}
\end{figure}

The experiment was performed on RT-1 device.
Figure\,\ref{f:schematic} shows the layout of the device, together with associated plasma images  (soft X-ray image and an electron density distribution reconstructed from interferometry). 
Plasma is produced by electron cyclotron resonance heating (ECH) with an 8.2\,GHz microwave (maximum power of 50 kW) and a discharge duration of approximately 1\,s.
The plasma contains high-temperature ($10\sim 50$ keV) electrons as well as low temperature ($\lesssim 100$ eV) electrons; typically, the former occupies about half of the total population\,\cite{Saitoh2011}.
Electrons are confined in a radially elongated and vertically thin region, resembling planetary radiation belts\,~\cite{Saitoh2014,Saitoh2015}; see Fig.\,\ref{f:schematic}(a).
From quasi-neutrality, ion density distribution is also considered to be peaked on the equatorial plane ($z=0$). 
In the inward diffusion process, a concomitant heating mechanism must be in effect to produce high-energy electrons.
However, electrons are also heated by ECH (cyclotron resonance occurs just in the vicinity of the levitated ring magnet).
It is, therefore, difficult to separate the betatron and Fermi accelerations in the total energy balance of electrons.
As a viable alternatives, ions may thus be used as an appropriate target for analyzing potential heating mechanisms.

Spatial distribution of ion temperature was measured by two sets of Doppler spectroscopies;
one scanned the plasma on a horizontal plane, and the other on a vertical plane (see Fig.\,\ref{f:schematic}).
The lines of sight of the former ranged from $r=375$\,mm to 878\,mm, and those of the latter 
ranged from $r=375$\,mm to 808\,mm on the $z=0$ axis ($r$ and $z$ are the radial and vertical coordinates, respectively, $r=375$\,mm is the surface of the superconducting magnet, $r=1000$\,mm is the vacuum chamber wall, and $z=0$ is the mid plane of the magnetic dipole).
He\,II transition (468.565\,nm) was used to measure the Doppler broadening.  
The horizontal chords measured the perpendicular temperature ($T_\perp$) with respect to the ambient magnetic field, 
while the vertical chords measured the mixture of $T_\perp$ and the parallel temperature ($T_{||}$).
The local temperature profiles of $T_\perp$ and $T_{||}$ were reconstructed from the line integrated spectrum data.
(see Appendix for the detailed algorithm).

\section{Observation of ion temperature anisotropy}\label{s:observation}
\begin{figure}[htpb]
	\begin{center}
		\includegraphics*[width=0.5\textwidth]{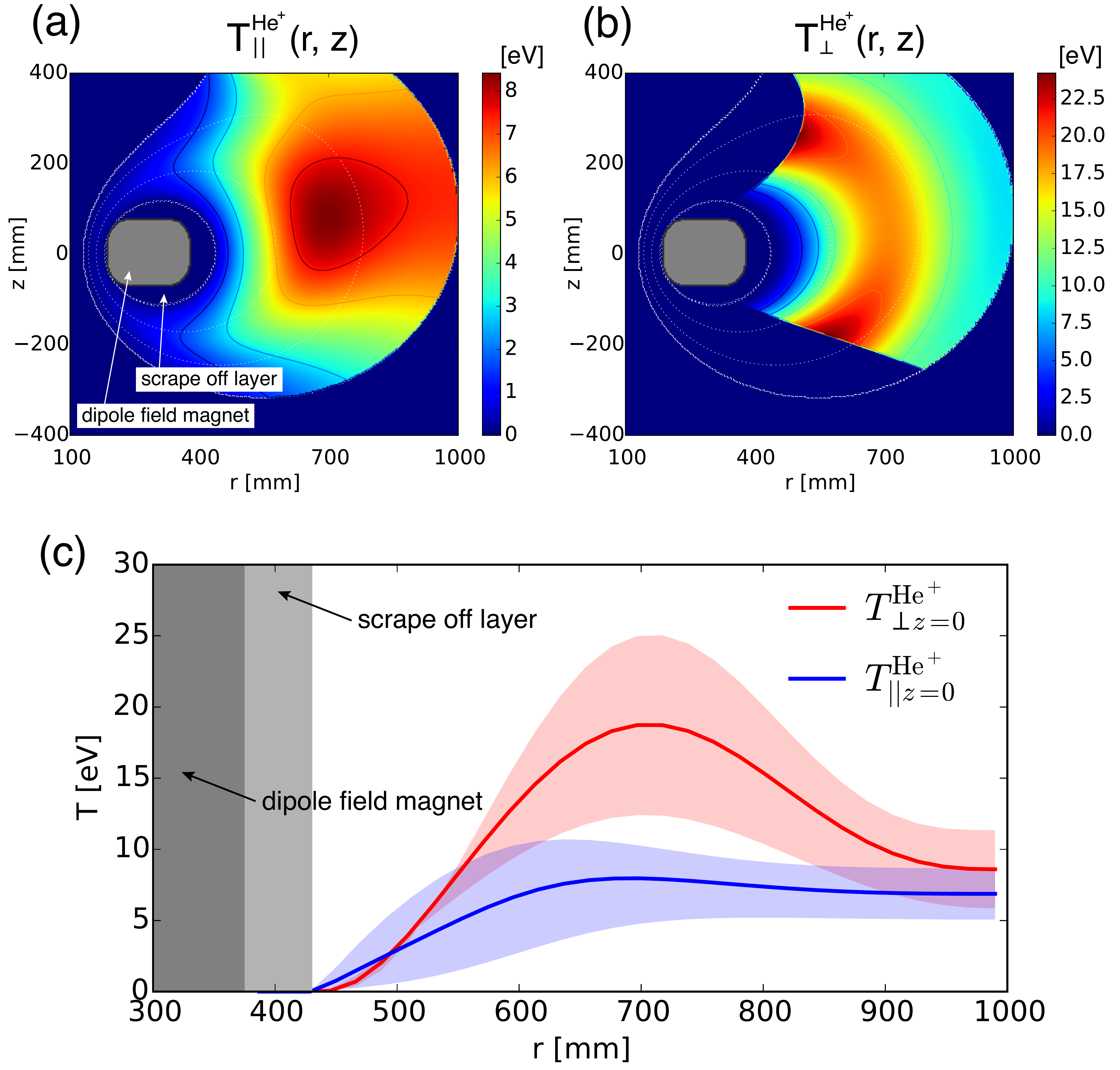}
	\end{center}
	\caption{ 
						Two dimensional profiles of (a) parallel temperature ($T_{||}$) and (b) perpendicular temperature ($T_\perp$) of He$^{+}$ ions.
						$T_\perp$ is plotted in a region where the mirror ratio (the local magnetic field normalized by the
						minimum magnetic field along each field line) is less than 2.
						(c) The radial temperature profile on the equatorial plane ($z=0$).
						The translucent regions are the error bands estimated by the covariance of the reconstruction fitting.
					}
	\label{f:profile}
\end{figure}

Figure\,\ref{f:profile}(a) and \ref{f:profile}(b) shows the two dimensional profiles of $T_{||}$ and $T_\perp$ of the He$^{+}$ ion in helium plasma.
Because of the mirror effect, $T_{||}$ decreases and $T_\perp$ increases with closer distance to the magnetic poles.
The radial temperature profile of the He$^{+}$ ion on the equatorial plane is plotted in Fig.\,\ref{f:profile}(c), showing an evident difference between $T_\perp$ and $T_{||}$.

\begin{figure}[htpb]
	\begin{center}
		\includegraphics*[width=0.5\textwidth]{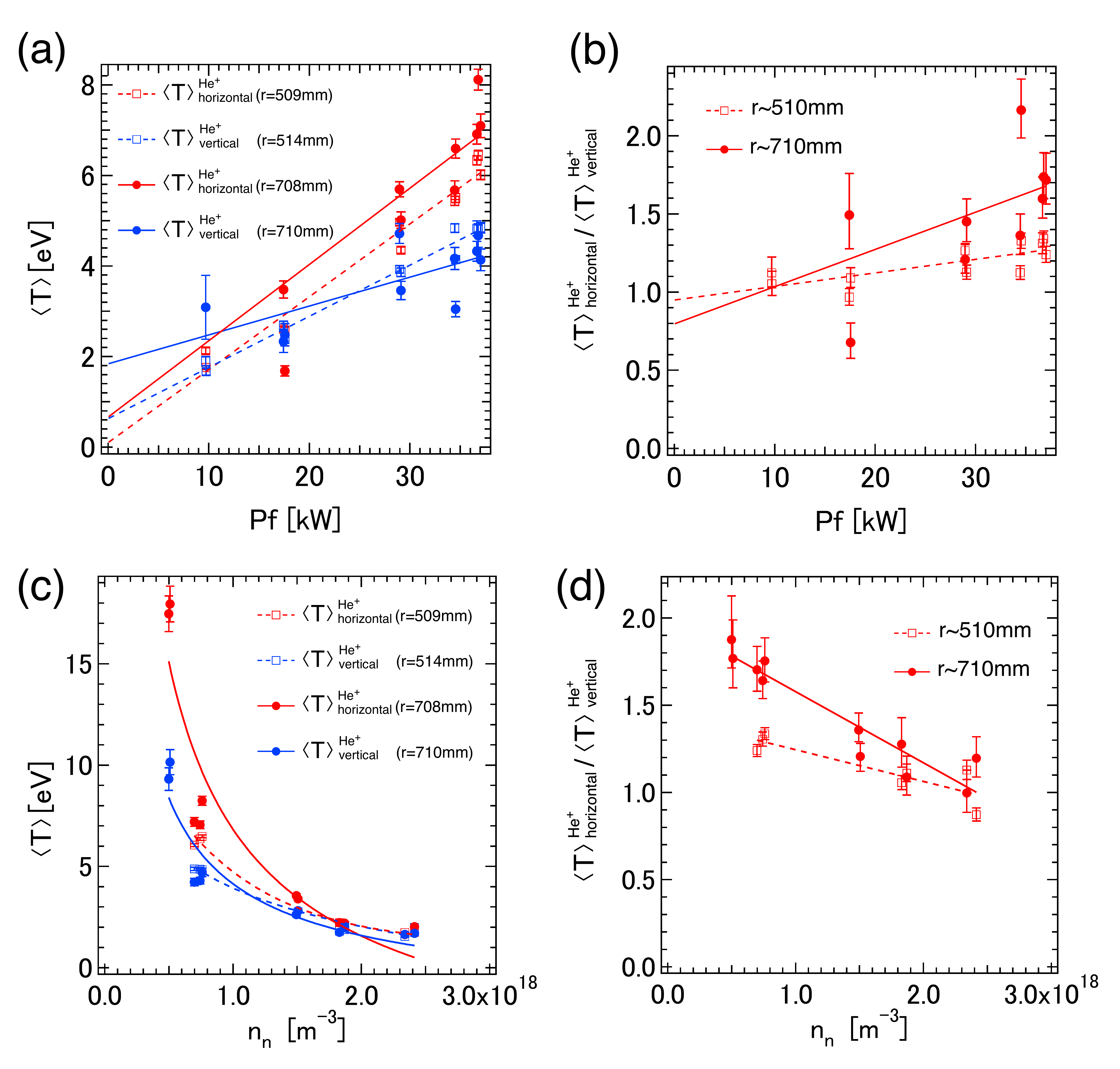}
	\end{center}
	\caption{ 
						Dependence of
						(a) the line averaged temperature of He$^{+}$ and (b) the anisotropy on the ECH power,
						and the dependence of (c) the line averaged temperature and (d) the anisotropy on the neutral helium gas density.
					}
	\label{f:dependence}
\end{figure}

Figure\,\ref{f:dependence} shows the dependencies of line averaged ion temperatures and anisotropy on ECH power and neutral helium particle density at two different radial points ($r\sim$510\,mm and $r\sim$710\,mm).
The neutral helium particle density is estimated as $n_\mr{n} = p/k_B T - n_\mr{e}$ where $p$ is the filling gas pressure, $k_B$ is the Boltzmann constant, $T = 300$\,K is the room temperature and $n_\mr{e}$ is the electron density measured by the interferometers. 
From Fig.\,\ref{f:dependence}(a) and \ref{f:dependence}(b), both the temperatures and anisotropy increase as ECH power increases.
From Fig.\,\ref{f:dependence}(c) and \ref{f:dependence}(d) both the temperatures and anisotropy decreases as the neutral helium particle density increases.

\section{Energy transport model including betatron acceleration}\label{s:model}
In order to distinguish the effect of betatron acceleration as the preferential heating mechanism for $T_\perp$,
the energy balance in the He$^{+}$ ions was examined.
By conservation of the magnetic moment, 
$T_{\perp}$ must increase upon displacement towards the central region as the magnetic field strength increases.
From conservation of the first adiabatic invariant, it may be estimated that $\Delta T_\perp/T_\perp = \Delta B/B$, where $B$ is the magnetic field strength.
Let $\Delta r$ denotes the radial displacement by an infinitesimal time interval $\Delta t$.
The heating power by betatron acceleration is given by
\begin{eqnarray}
	P_\mr{betatron} \sim \f{\Delta T_\perp}{\Delta t} = T_\perp V_r\l|\dd{\ln B}{r}\ri|,
\label{e:Pbetatron}
\end{eqnarray}
where $V_r = \Delta r/\Delta t$ is the speed of inward diffusion.

A close parallel mechanism is the Fermi acceleration which heats $T_\parallel$ via an increase in the bounce frequency.
In a point-dipole magnetic field (which may approximate a planetary magnetic field),
it is known that betatron acceleration yields much stronger heating than Fermi acceleration\,~\cite{Dessler}.
In the magnetic field of RT-1, on the other hand, both of them may yield almost same powers if the first and second actions are equally conserved.
But if the constancy of the second action is broken upon inward diffusion, only the betatron acceleration will be in effect.
The relative importance of these heating mechanisms depends on the driving mechanism of the inward diffusion.

\begin{figure}[htpb]
	\begin{center}
		\includegraphics*[width=0.5\textwidth]{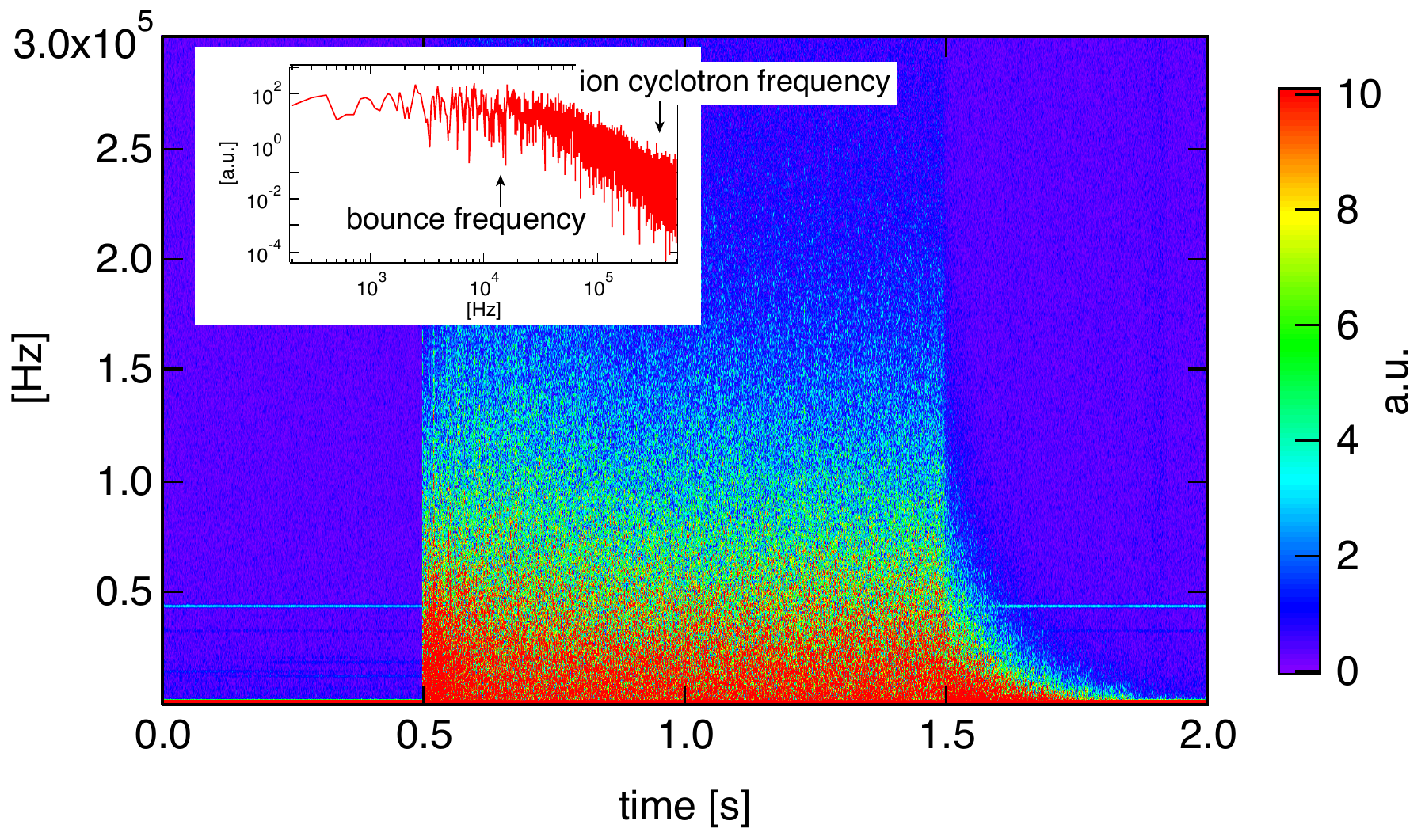}
	\end{center}
	\caption{
					Time evolution of the frequency spectrum measured by the reflectometer for the total lifetime of the plasma (ECH power is injected from 0.5\,s through 1.5\,s). 
					The density cutoff resides at $r \sim 640$\,mm.
					The inset is the clip between $t=1.001584$\,s and $t=1.063193$\,s.
					The fluctuation has a flat spectrum up to around 30 kHz (which ranges beyond the bounce frequency).
					However, the fluctuations in the range of the ion cyclotron frequency are very weak.
					}
	\label{f:reflectometer}
\end{figure}

Figure\,\ref{f:reflectometer} shows the frequency spectrum of the density fluctuations detected by a reflectometer.  
A flat spectrum extending up to around $10$\,kHz was observed.
The bounce frequency of a He$^{+}$ ion is typically around 15\,kHz (assuming $T_\parallel\sim 10$\,eV).
On the other hand, the ion cyclotron frequency is typically 360\,kHz (for $T_{\perp} = 20$\,eV and $B = 150$\,G); thus, the betatron acceleration is deemed to be ideally in effect.
Hence, it was concluded that in this instance, the Fermi acceleration is at most very weak in comparison with the betatron acceleration.

Other competitive processes are 
(1) the thermal equilibration of He$^{+}$ ions and the cold component of electrons,
(2) the thermal equilibration of He$^{+}$ ions and He$^{+}$ ions immediately after ionization from neutral helium particles which have isotropic temperature $T_0$,
(3) the loss of He$^{+}$ ions by charge exchange with neutral particles
and (4) the isotropization of $T_\perp$ and $T_{||}$.
The effect of hot electrons may be neglected because the thermal equilibration time is on the order of $10^3$\,s, which is considerably longer than the lifetime of the plasma.
In this instance, no assumption is made of any `anomalous transport'; thus, the only energy loss mechanism at play is charge exchange loss.

Taking into account the aforementioned processes,
a one-dimensional model of energy transport on the radial coordinate $r$ may be considered:
\begin{eqnarray}
	\label{e:dTperpdt}
	\dd{T_\perp}{t} &=& \frac{T_\mr{e} - T_\perp}{\tau_\mr{ei}} + \frac{T_\mr{0} - T_\perp}{\tau_\mr{ii}} - \frac{T_\perp - T_{||}}{\tau_\mr{iso}} - \frac{T_\perp}{\tau_{\mathrm{cx}}} \nonumber\\
									& & + P_\mr{betatron} , \\
																				 \nonumber\\
	\label{e:dTparadt}
	\dd{}{t}\l(\f{T_{||}}{2}\ri) &=& \frac{T_\mr{e} - T_{||}}{2\tau_\mr{ei}} + \frac{T_\mr{0} - T_{||}}{2\tau_\mr{ii}} + \frac{T_\perp - T_{||}}{\tau_\mr{iso}} - \frac{T_{||}}{2\tau_{\mathrm{cx}}}, \nonumber\\
\end{eqnarray}
where $\rmd/\rmd t$ is the temporal derivative to be evaluated on each volume element
co-moving with the plasma
on a Lagrangian coordinate;
for a stationary state, we may evaluate $\rmd/\rmd t=V_r \rmd/\rmd r$ with an inward diffusion speed $V_r$.
All coefficients of equations (\ref{e:dTperpdt}) and (\ref{e:dTparadt}) are evaluated by experimentally measured local plasma parameters.

\begin{figure}[htpb]
	\begin{center}
		\includegraphics*[width=0.5\textwidth]{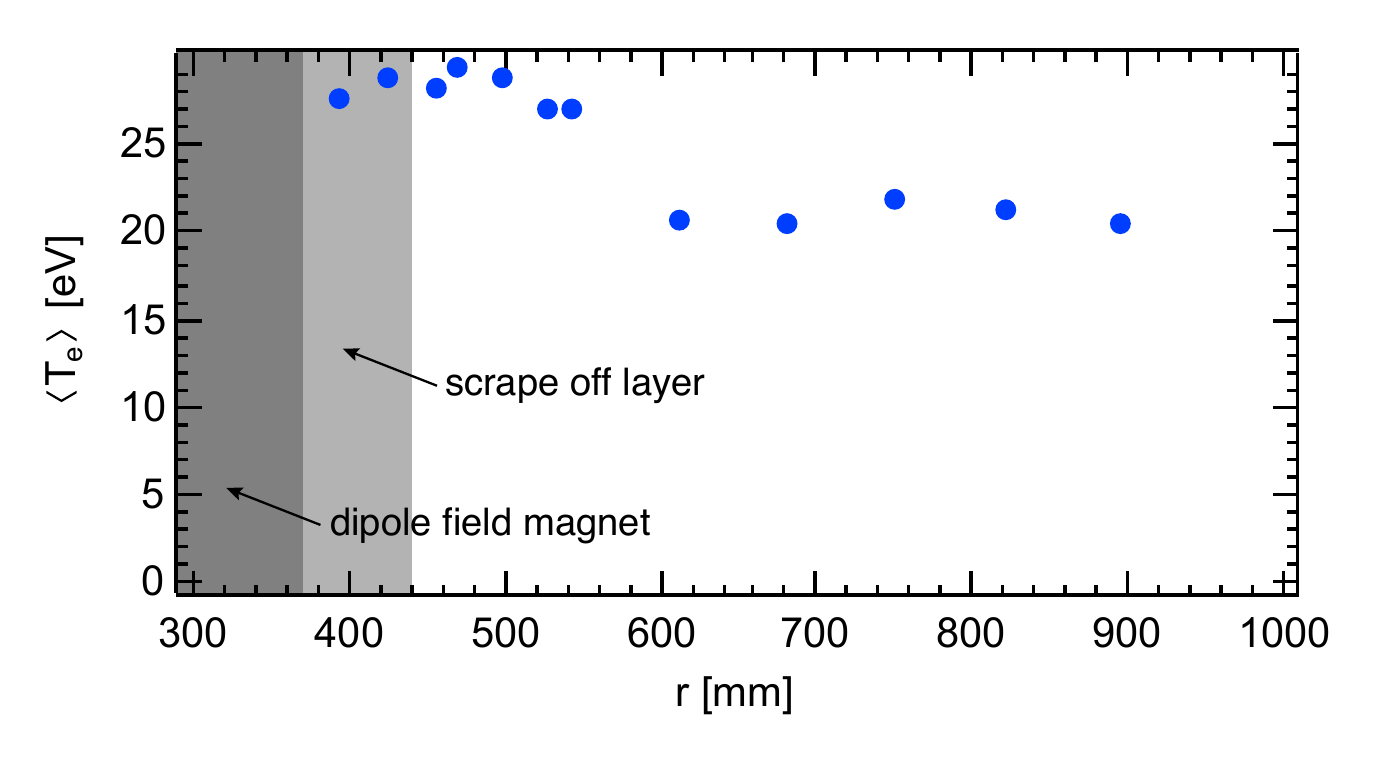}
	\end{center}
	\caption{
					The spatial profile of $T_\mr{e}$ measured by line integrated He~I line ratios (728.1/706.5 and 667.8/728.1 nm).
					The plasma condition is different from that of Fig.~\ref{f:profile}.
					}
	\label{f:Te}
\end{figure}

The electron-ion thermal equilibration time ($\tau_\mr{ei}$) is a function of the ion temperature ($\sim 10$ eV), the cold electron density ($n_\mr{e} \sim 10^{17}~\mathrm{m}^{-3}$ measured by the interferometers) and the cold electron temperature $T_\mr{e}$~\cite{Huba}.
Figure~\ref{f:Te} shows the spatial profile of $T_\mr{e}$ measured by line integrated He~I line ratios (728.1/706.5 and 667.8/728.1 nm)~\cite{Schweer} for different plasma condition from that of Fig.~\ref{f:profile}.
Unlike the ion temperature profile, $T_\mr{e}$ is higher near to the dipole field magnet.
We only have the single spatial point measurement $T_\mr{e} = 30.3$\,eV at $r=795$\,mm for the same condition as Fig.~\ref{f:profile}.
Therefore, we infer that the cold electron temperature is distributed as $30 \lesssim T_\mr{e} \lesssim 40$\,eV in the plasma for the present study.
Finally, $\tau_\mr{ei}$ was determined to be $\sim 0.1$\,s.
The thermal equilibration time among He$^+$ ions ($\tau_\mr{ii}$) is estimated as $\sim 1$\,ms by the ion temperature and density ($\simeq n_\mr{e}$).
The ion isotropization time ($\tau_\mr{iso}$) is on the order of 1\,ms, as estimated by the ion temperatures ($T_\perp \sim 20$\,eV and $T_{||} \sim 10$\,eV)~\cite{Huba}.

\begin{figure}[htpb]
	\begin{center}
		\includegraphics*[width=0.5\textwidth]{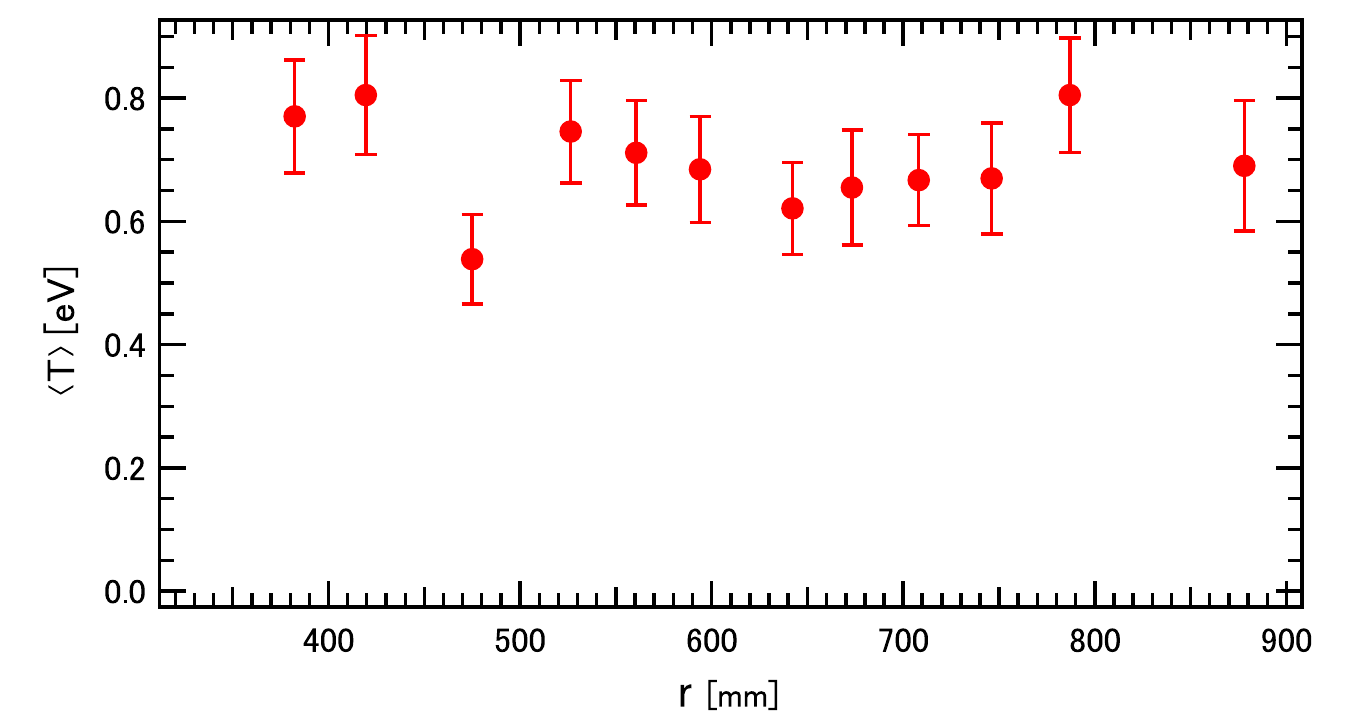}
	\end{center}
	\caption{
					Radial profile of the line averaged temperature of the neutral helium particles evaluated from the Doppler broadening of the He\,I line spectrum (471.315\,nm) on the horizontal line of sight.
					}
	\label{f:THe0}
\end{figure}

\begin{figure}[htpb]
	\begin{center}
		\includegraphics*[width=0.5\textwidth]{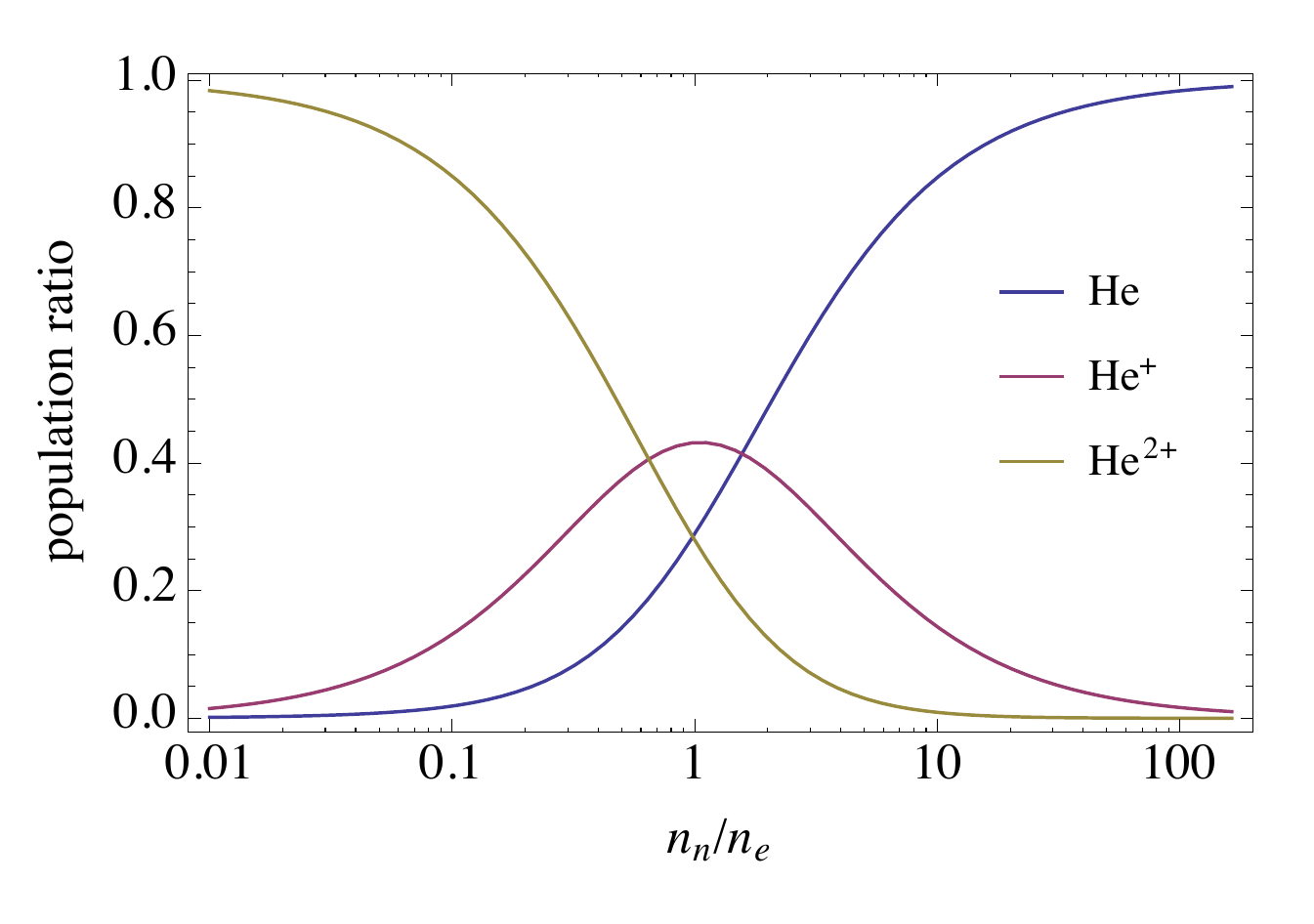}
	\end{center}
	\caption{
					Ionization ratio of helium as a function of neutral particle density, obtained by solving the rate equation of ionization, recombination and charge exchange processes.
					The electron temperature was set to 30\,eV from the measurement.
					}
	\label{f:ionization ratio}
\end{figure}

Let us estimate neutral helium gas density profile to evaluate the time constant $\tau_{\mathrm{cx}}$ which scales the time of charge transfer between the He$^{+}$ ion and the neutral helium gas.
The mean free path of the neutral particle is $\lambda = v_\mr{n}/n_\mr{e}\l<\sigma v\ri>_\mr{ion}$, where $v_\mr{n}$ is the thermal velocity of the neutral gas and $\l<\sigma v\ri>_\mr{ion}$ is the ionization rate coefficient.
The neutral helium temperature was found to be almost spatially homogeneous with 0.7\,eV by the Doppler broadening of the He\,I (471.315\,nm) line as shown in Fig.\,\ref{f:THe0}.
Using $n_\mr{e}$ of the shot in Fig.\,\ref{f:profile} and $30 \lesssim T_\mr{e} \lesssim 40$\,eV, $\lambda$ is evaluated as $\sim 1$\,m near the dipole field magnet and $\sim 10$\,m at the chamber wall, implying that the neutral gas density is spatially flat.
In addition, the absolute value of the neutral helium density must be evaluated.
We solved the rate equation of ionization, recombination and charge exchange processes of helium, varying neutral helium gas density.
The stationary solutions are shown in Fig.\,\ref{f:ionization ratio}.
Since adequate light emission was observed from both the He\,I and He\,II lines, the neutral gas density was determined to be roughly in the range of $0.1n_\mr{e} \lesssim n_\mr{n} \lesssim 10n_\mr{e}$.
If $n_\mr{n}$ is too small, He$^{2+}$ dominates and He\,I line would be scarce. 
Or if $n_\mr{n}$ is too large, He\,II line would disappear 
(since He$^{2+}$ does not emit light, we have yet to determine how much of He$^{2+}$ is present).
Therefore, the neutral helium gas density is homogeneously $n_\mr{n} \sim 10^{17}\,\mr{m^{-3}}$.
$\tau_{\mathrm{cx}}$ was determined to be $\sim 1$\,ms.

\begin{figure}[htpb]
	\begin{center}
		\includegraphics*[width=0.5\textwidth]{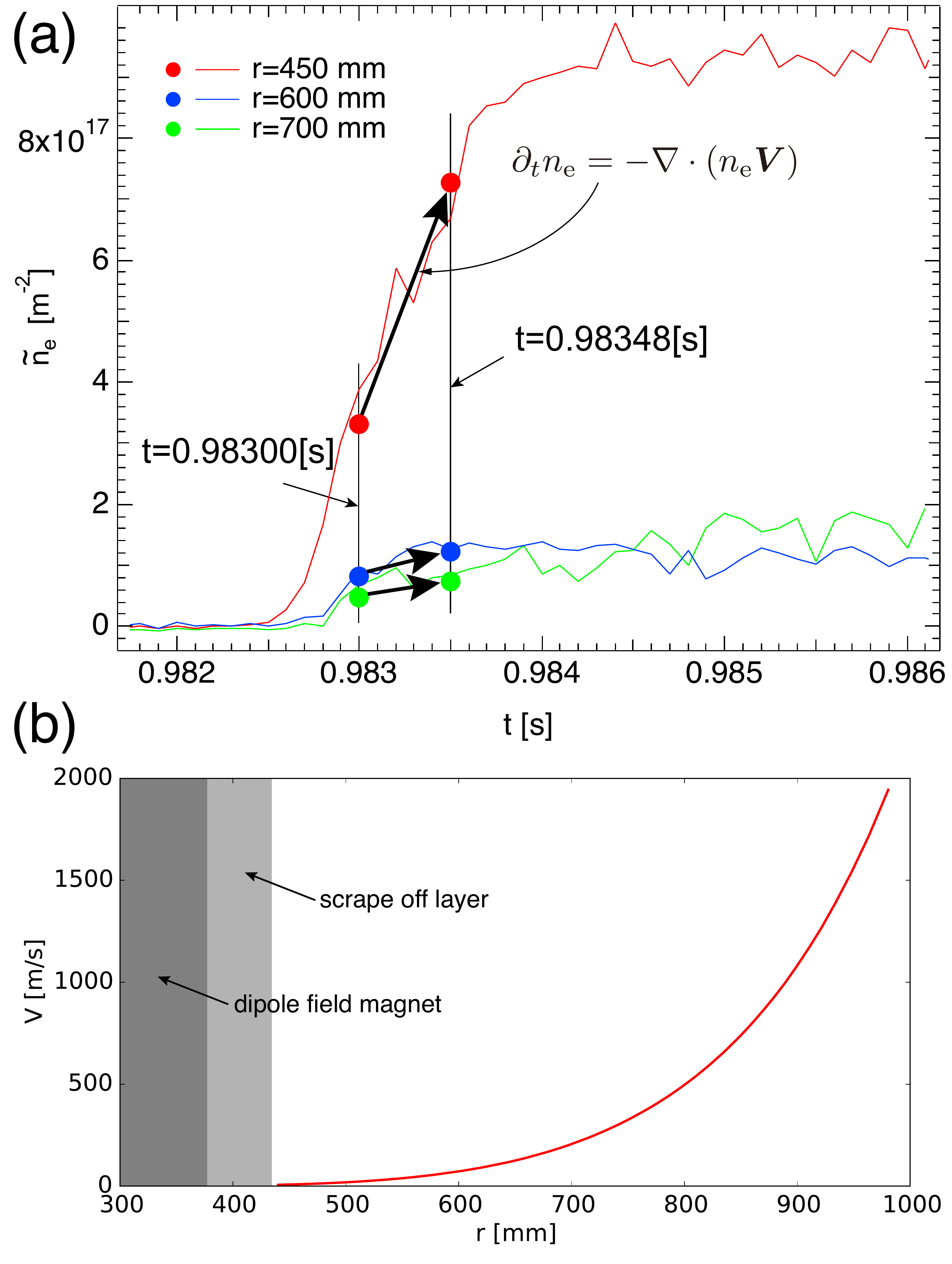}
	\end{center}
	\caption{
					(a) Time evolution of the line integrated electron density measured by the interferometers at each chords around the onset of the plasma discharge.	
					The bullets at $t=0.98348$\,s are the solution of (\ref{e:continuity}) with the parameters $(V_{\psi0},\, \alpha) = (3000,\, -7)$ and the initial condition as the bullets at $t=0.98300$\,s.
					(b) Radial distribution of the estimated inward diffusion speed $V_r$ on the equatorial plane.
					}
	\label{f:onset}
\end{figure}

The inward diffusion speed $V_r$ is required to evaluate $P_\mr{betatron}$.
This was measured in the initial formation phase.
At the beginning of the discharge, the density has a broad distribution, but subsequently it begins to concentrate into the central region, thereby creating a clump of particles.
Figure\,\ref{f:onset}(a) shows the time evolution of the line integrated electron density measured by the three chord interferometers around the onset of the plasma discharge.
Two dimensional density profile of $n_\mr{e}$ is reconstructed at the two different time points $t=0.98300$\,s and $t=0.98348$\,s.
The electron density evolves in accordance with the continuity equation with the inward diffusion velocity $\bm{V}$:
\begin{equation}
	\pp{n_\mr{e}}{t} + \f{1}{r}\pp{}{t}\l( rn_\mr{e}V_r \ri) + \pp{}{z}\l( n_\mr{e}V_z \ri) = 0.
\label{e:continuity}
\end{equation}
We assume $\bm{V}$ as 
\begin{equation*}
	\bm{V} = V_{\psi0}\l(\f{\psi}{\psi_1}\ri)^\alpha\f{\nabla\psi}{|\nabla\psi|}.
\end{equation*}
The parameters $V_{\psi0}$ and $\alpha$ are optimized so that the solution of (\ref{e:continuity}) with the initial condition as $n_\mr{e}$ at $t=0.98300$\,s in Fig.\,\ref{f:onset} becomes closer to $n_\mr{e}$ at $t=0.98348$\,s.
Choosing appropriate parameters, the density evolution is replicated well as shown in Fig.\,\ref{f:onset}(a), thus the speed $V_r$ that transports the particles is deduced.
In the stationary phase, we do not have a direct measurement of $V_r$, but we assume
that the same estimate applies.  
Figure\,\ref{f:onset}(b) shows optimized $V_r$ on the equatorial plane which will be used to evaluate $P_\mr{betatron}$.

\section{Numerical siumlation of the energy transport model}\label{s:simulation}
\begin{figure}[htpb]
	\begin{center}
		\includegraphics*[width=0.5\textwidth]{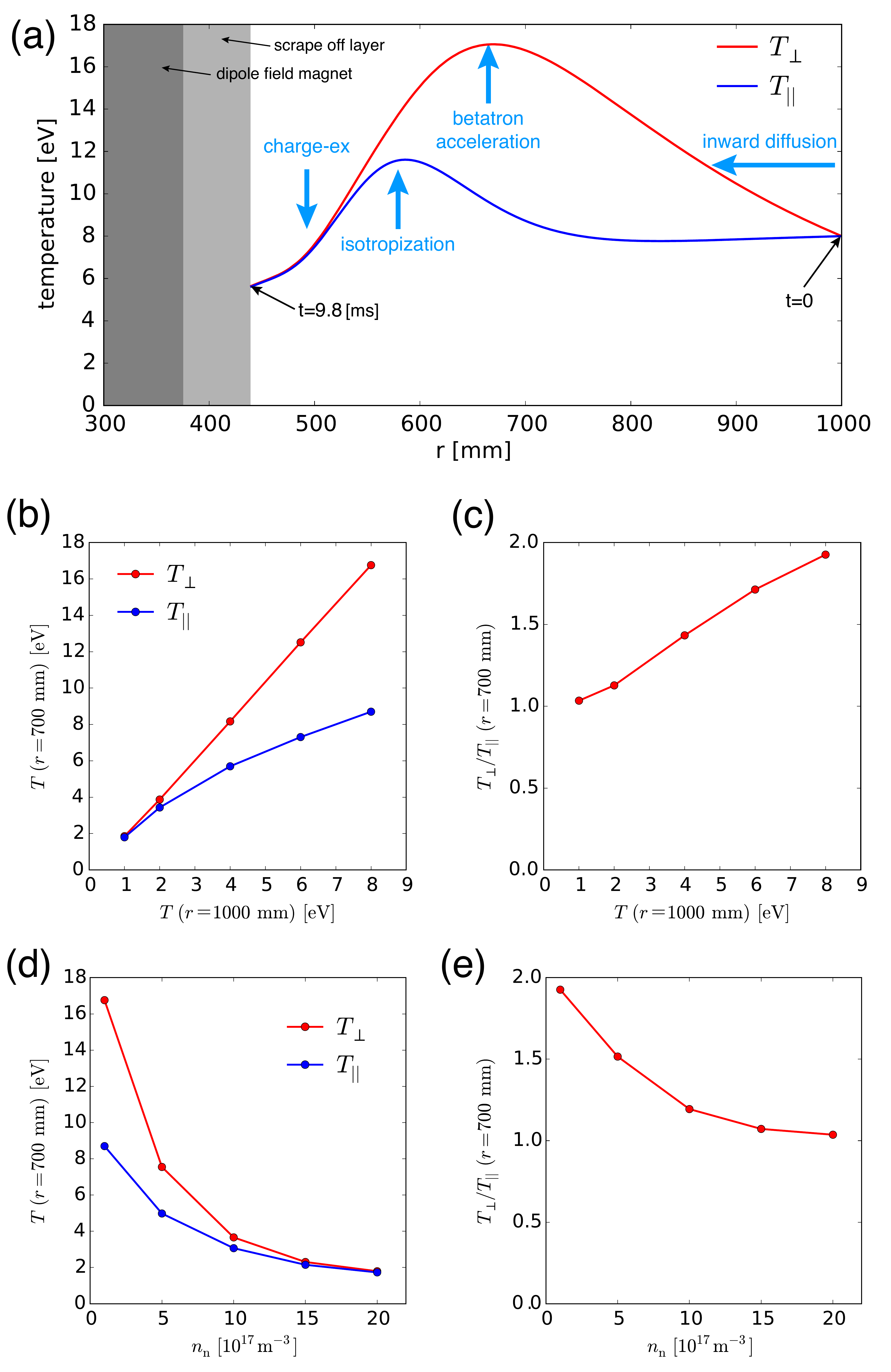}
	\end{center}
	\caption{
					(a) The radial profiles of $T_{\perp}$ and $T_{||}$ calculated by the energy transport model (\ref{e:dTperpdt})--(\ref{e:dTparadt}). 
					The result is compared with the experimental observation shown in Fig.\,\ref{f:profile}(c).
					Dependence of
					(b) the ion temperature and (c) the anisotropy on the initial temperature at $r=1000$\,mm,
					and the dependence of (d) the ion temperature and (e) the anisotropy on the neutral helium gas density.
					The temperatures and anisotropy are evaluated at $r=700$\,mm.
					The results are compared with the experimental scaling shown in Fig.\,\ref{f:dependence}.
					}
	\label{f:simulation}
\end{figure}

Equations (\ref{e:dTperpdt}) and (\ref{e:dTparadt}) are both ultimately solved with the aforementioned parameters.
An initial condition at $r=1000$ mm (edge of the plasma) is assumed;
using the measurement, $T_{\perp} = T_{||} = 8$ eV.
This initial temperature may be equivalent to the temperature of ions immediately after ionization from neutral helium particles.
Therefore we set $T_0 = 8$\,eV.
The solution is shown in Fig.\,\ref{f:simulation}(a) and it turns out that the model explains the experimental profiles presented in Fig.\,\ref{f:profile}(c).
In the first half of the inward transport ($1000 > r > 700$\,mm), the betatron acceleration is stronger than the charge exchange loss because of the sufficiently fast $V_r$.  
This results in the increase in $T_\perp$.
In the last half ($700 > r > 375$\,mm) where $V_r$ is rather smaller, on the other hand, the charge exchange loss overcomes the betatron acceleration.
Thus, $T_\perp$ decreases.
Since $\tau_\mr{iso}$ is smaller than the inward diffusion time scale in this region, the isotropization also decreases $T_\perp$, and at the same time, increases $T_{||}$
Here the exact value of the neutral gas density (which has a homogeneous distribution in the plasma) is tuned to $1\times 10^{17}\,\mr{m^{-3}}$ to match the peak value of $T_\perp$.

Figure\,\ref{f:simulation}(b)-\ref{f:simulation}(e) illustrate the dependencies of temperatures and anisotropy calculated from our model on the initial temperature at $r=1000$\,mm and the neutral helium particle density.
The results are consistent with the experimental scaling in Fig.\,\ref{f:dependence}.
The increase in the ECH power implies an increase in the initial ion temperature through the relaxation between cold electrons; then $P_\mr{betatron}$ becomes larger, resulting in larger $T_\perp$ and anisotropy.
The increase in $T_{||}$ is given by the isotropization.
The decrease in the ion temperature with the increase in neutral helium gas density is consistent with the charge exchange loss term in (\ref{e:dTperpdt}) and (\ref{e:dTparadt}). 
Since $P_\mr{betatron}$ is independent of the neutral gas density, an increase in the charge exchange loss results in small anisotropy.
Therefore all the dependencies were found to be consistent with the assessed model underpinning this study.

In conclusion, the anisotropic ion temperature observed in the laboratory radiation belt is experimental proof of the
betatron acceleration concomitant with inward radial diffusion of particles.

\begin{acknowledgments}
We appreciate the helpful comments and suggestions by Dr. Shinichiro Kado.  
This work was supported by JSPS KAKENHI Grant No. 23224014. 
\end{acknowledgments}

\clearpage
\appendix
\section{Algorithm for the reconstruction of the local profile}

To reconstruct the local temperature profile from the line averaged measured data, the following procedure was used.
The local profile on the equatorial plane ($z=0$) is reconstructed from the spectrum data obtained via the horizontal chord.
From the axisymmetric property, the spectrum on the equatorial plane can be written as $f_\mr{eq}(\lambda,\; \psi(r,\,0))$
where $\lambda$ is the wavelength, $\psi(r,\,z)$ is the inverse of magnetic flux and $r$ and $z$ are the radial and vertical coordinates, respectively.
The innermost flux surface contacting the dipole field magnet is denoted as $\psi_0$, and the outermost flux contacting the vacuum chamber at $r = 1000\,\mr{mm}$ is denoted as $\psi_1$.
Hereafter, the subscript $\mr{eq}$ means the value on the equatorial plane.
The line integrated spectrum $g_\mr{eq}(\lambda,\; y)$ is thus calculated from the local spectrum as
\begin{equation}	
	g_\mr{eq}(\lambda,\, y) = 2\int_0^{1} f_\mr{eq}(\lambda,\, \psi_\mr{eq}(\sqrt{x^2 + y^2}))\rmd x,
\label{e:toridal transform}
\end{equation}
where $x$ denotes the auxiliary coordinate along the line of sight.
We assume  
\begin{widetext}
\begin{eqnarray*}
	f_\mr{eq}(\lambda,\; \psi_\mr{eq}(r)) = 
	\left\{
	\begin{array}{cc}
	 \ds {\mathcal A}_\mr{eq}(\psi_\mr{eq}(r)) \exp\l[ -\l( \f{\lambda - \lambda_0 - {\mathcal B}_\mr{eq}(\psi_\mr{eq}(r))}{{\mathcal C}_\mr{eq}(\psi_\mr{eq}(r))}  \ri)^2 \ri] \quad (\psi_0 \le \psi_\mr{eq}(r) \le \psi_1)\\
	 \\
	 0  \quad \quad \quad \quad \quad \quad\quad (\psi_1 \le \psi_\mr{eq}(r))\\
	\end{array}
	\right.
	,
\end{eqnarray*}
\end{widetext}
where $\lambda_0$ is the central wavelength.
The intensity ${\mathcal A}_\mr{eq}(\psi_\mr{eq}(r))$ and the shift ${\mathcal B}_\mr{eq}(\psi_\mr{eq}(r))$ are assumed to be 
\begin{eqnarray*}
	{\mathcal A}_\mr{eq}(\psi_\mr{eq}(r)) &=& a_1\f{(\psi_\mr{eq}/\psi_0 - 1)^{a_2}(1 - \psi_\mr{eq}/\psi_1)^{a_3}}{(\psi^*/\psi_0 - 1)^{a_2}(1 - \psi^*/\psi_1)^{a_3}} \\ 
	\\
	& & \quad  \quad \quad \quad \quad \quad \l( \psi^* = \f{a_3 \psi_0 + a_2 \psi_1}{a_2 + a_3} \ri) \\
	\\
	{\mathcal B}_\mr{eq}(\psi_\mr{eq}(r)) &=& b_1 + b_2\psi_\mr{eq} + b_3\psi_\mr{eq}^2,
\end{eqnarray*}
with the perpendicular temperature on the equatorial plane assumed to be
\begin{eqnarray}
	T_{\perp \mr{eq}}(\psi_\mr{eq}) &=& (t_1 - 0.5t_4)\f{(\psi_\mr{eq}/\psi_0 - 1)^{t_2}(1 - \psi_\mr{eq}/\psi_1)^{t_3}}{(\psi^\dagger/\psi_0 - 1)^{t_2}(1 - \psi^\dagger/\psi_1)^{t_3}} \nonumber\\
	\nonumber\\ 
	& & \quad \quad + 0.5t_4\l[1+\mr{erf}\l(\f{\psi - \psi^\dagger}{0.2(\psi_1 - \psi_0)}\ri) \ri] \nonumber\\
	\nonumber\\ 
	& &  \quad \quad \quad \quad \quad \quad \l( \psi^\dagger = \f{t_3 \psi_0 + t_2 \psi_1}{t_2 + t_3} \ri).
	\label{e: T_perp0}
\end{eqnarray}
The Doppler broadening ${\mathcal C}_\mr{eq}(\psi_\mr{eq})$ is determined by $T_{\perp\mr{eq}}(\psi_\mr{eq})$.
The functions ${\mathcal A}_\mr{eq}, {\mathcal B}_\mr{eq}$ and $T_{\perp \mr{eq}}$ indicate that the spectral emission is zero at both boundaries, the temperature is zero at $\psi = \psi_0$ and $t_1$ at $\psi = \psi_1$, and the toroidal flow speed may be finite at both boundaries.
The observed spectrum data set for the horizontal chord is written as $(\lambda_i,\, y_j,\, g_{\mr{eq}ij})$.
The parameters $(a_1 ,\,a_2 ,\,a_3 ,\,b_1 ,\,b_2 ,\,b_3 ,\,t_1 ,\,t_2 ,\,t_3 ,\,t_4)$ are optimized so as to minimize $|g_{\mr{eq}\,ij} - g_\mr{eq}(\lambda_i,\, y_j)|$ for each $\lambda_i$ and $y_j$.

Next we build a vertical reconstruction algorithm.
Using ${\mathcal A}_\mr{eq}$ and $T_\mr{eq}$ obtained by the horizontal reconstruction, a two-dimensional profile is reconstructed.
The position of the collimator is denoted as $(r_\mr{c},\, z_\mr{c})$ (Fig.\,\ref{f:schematic}(a)).
The line of sight passing the equatorial point $(y,\, 0)$ is given as
\begin{equation}
	r = \f{r_\mr{c} - y}{z_\mr{c}}(z - z_\mr{c}) + r_\mr{c}.
\label{e:polidal L}
\end{equation}
The line integrated spectrum $g(\lambda,\, y)$ is calculated from the local spectrum $f(\lambda,\, r,\, z)$ as
\begin{equation}
	g(\lambda,\, y) = \int_{-z_\mr{c}}^{z_\mr{c}}  f(\lambda,\, r(z),\, z) \sqrt{1 + \l( \dd{r}{z} \ri)^2}\rmd z.
\label{e:vertical transformation}
\end{equation}
In the same manner as the horizontal reconstruction, the local spectrum is assumed to be
\begin{eqnarray*}
	f(\lambda,\, r,\, z) = 
	\left\{
	\begin{array}{cc}
	 \ds {\mathcal A}(r,\, z) \exp\l[ -\l( \f{\lambda - \lambda_0}{{\mathcal C}(r,\, z)}  \ri)^2 \ri] & (\psi_0 \le \psi \le \psi_1) \\
	 \\
	 0  & (\psi_1 \le \psi)\\
	\end{array}
	\right.
	,
\end{eqnarray*}
In the vertical chord, the Doppler shift can be ignored because of the absence of the mean poloidal flow in the dipole magnetic field configuration.
We use the magnetic field coordinate to treat the vertical direction, and assume ${\mathcal A}(r,\, z)$ to be the power function of the magnetic field strength $B$.
\begin{eqnarray}
	{\mathcal A}(r,\, B(r,\, z)) = {\mathcal A}_\mr{eq}(\psi_\mr{eq})\l[ \f{B(r,\, z)}{B_\mr{eq}} \ri]^{-\tilde{a}}
\label{e: A}
\end{eqnarray}
where $B_\mr{eq}$ is the magnetic field strength of the same field line on the equatorial plane.
In the $z \ne 0$ region, the effective temperature ($T_\mr{eff}$) causing the Doppler broadening is a combination of $T_\perp$ and $T_{||}$.
Defining the angle $\theta$ between the line of sight and the line normal to the magnetic field, we may write
\begin{equation}
	(T_\text{eff})^2 = (T_\perp \cos\theta)^2 + (T_{||} \sin\theta)^2.
\label{e: Tobs}
\end{equation}
In the same manner as (\ref{e: A}), we assume the perpendicular temperature profile as
\begin{equation}
	T_\perp(r,\, B(r,\, z)) = T_{\perp\mr{eq}}\l( \f{B}{B_\mr{eq}} \ri)^{\tilde{c}_1}.
\label{e: Tperp}
\end{equation}
While $T_{\perp\mr{eq}}$ is already determined by the horizontal reconstruction, we need the profile of $T_{||}$ on the equatorial plane.
We assume $T_{||\mr{eq}}$ in a similar manner as $T_{\perp\mr{eq}}$:  
\begin{eqnarray}
	T_{||\mr{eq}}(\psi_\mr{eq}(r)) &=& (\tilde{t}_1 - 0.5\tilde{t}_4)\f{(\psi_\mr{eq}/\psi_0 - 1)^{\tilde{t}_2}(1 - \psi_\mr{eq}/\psi_1)^{\tilde{t}_3}}{(\psi^\ddagger/\psi_0 - 1)^{\tilde{t}_2}(1 - \psi^\ddagger/\psi_1)^{\tilde{t}_3}} \nonumber\\
	& & \quad \quad + 0.5\tilde{t}_4\l[1+\mr{erf}\l(\f{\psi - \psi^\ddagger}{0.2(\psi_1 - \psi_0)}\ri) \ri] \nonumber\\
	\nonumber\\ 
	& & \quad \quad \quad \quad \quad \quad \l( \psi^\ddagger = \f{\tilde{t}_3 \psi_0 + \tilde{t}_2 \psi_1}{\tilde{t}_2 + \tilde{t}_3} \ri).
\label{e: T||0}
\end{eqnarray}
In the same manner as (\ref{e: A}) and (\ref{e: Tperp}), we assume
\begin{equation}
	T_{||}(r,\, B(r,\, z)) = T_{||\mr{eq}}\l( \f{B}{B_\mr{eq}} \ri)^{-\tilde{c}_2}.
\label{e: T||}
\end{equation}
We write the observed spectrum data set for the vertical chord as $(\lambda_i,\, y_j,\, g_{ij})$.
The parameters $(\tilde{a},\, \tilde{c}_1,\, \tilde{c}_2,\, \tilde{t}_1,\, \tilde{t}_2,\, \tilde{t}_3,\, \tilde{t}_4)$ are optimized to minimize $|g_{ij} - g(\lambda_i,\, y_j)|$ for each $\lambda_i$ and $y_j$.
A two-dimensional profile of ${\mathcal A}(r,\, z)$, $T_\perp(r,\, z)$ and $T_{||}(r,\, z)$ is then ultimately obtained.

\begin{figure}[htpb]
	\begin{center}
		\includegraphics*[width=0.5\textwidth]{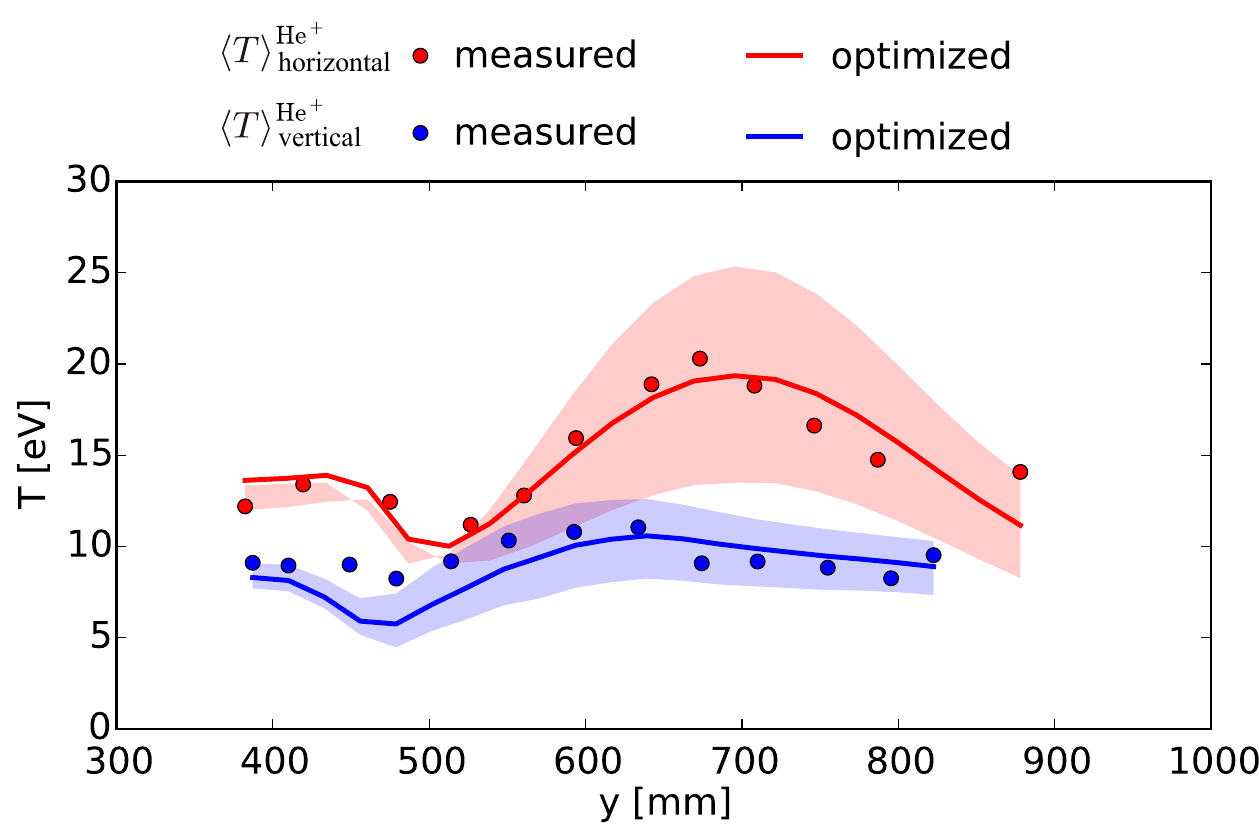}
	\end{center}
	\caption{
						Line averaged temperature profile to estimate the fitting precision.
						The bullets are the line averaged temperatures of He$^+$ estimated by the measured spectrum data.
						The solid lines are the line averaged temperatures estimated by the optimized $g$.
						The translucent regions are the error bands determined by multiplying a constant by the covariance of the reconstruction fitting.
					}
	\label{f:optimization}
\end{figure}

Figure\,\ref{f:optimization} shows the fitting precision of Fig.\,\ref{f:profile}(c).
The solid lines are the line averaged temperatures estimated by $g_\mr{eq}(\lambda,\, y)$ and $g(\lambda,\, y)$, respectively, with optimized parameters, while the bullets are the line averaged temperatures estimated by the measured data $g_{\mr{eq}\,ij}$ and $g_{ij}$, respectively. 
The translucent regions are the error bands determined by multiplying a constant by the covariance of the reconstruction fitting.
The corresponding error bands for the local profile are illustrated in Fig.~\ref{f:profile}(c).

\clearpage



\begin{thebibliography}{99}

\bibitem{Schulz}
M. Schulz and L. J. Lanzerotti,  
\textit{Particle Diffusion in the Radiation Belts},
(Springer, 1974).

\bibitem{Dessler}
A. J. Dessler, ed. 
\textit{Physics of the Jovian magnetosphere}., 
{\bf 3}. (Cambridge University Press, 2002).

\bibitem{Birmingham}
T. J. Birmingham, T. G. Northrop, and C.-G. F\"{a}lthammar, 
\textit{Charged Particle Diffusion by Violation of the Third Adiabatic Invariant},
Phys. Fluids {\bf 10}, 2389-2398 (1967)

\bibitem{Hasegawa}
A. Hasegawa,
\textit{Motion of a Charged Particle and Plasma Equilibrium in a Dipole Magnetic Field Can a Magnetic Field Trap a Charged Particle? Can a Magnetic Field Having Bad Curvature Trap a Plasma Stably?},
Phys. Scr. T {\bf 116}, 72 (2005).

\bibitem{YM2014}
Z. Yoshida and S. M. Mahajan, 
\textit{Self-organization in foliated phase space: construction of a scale hierarchy by adiabatic invariants of magnetized particles},
Prog. Theor. Exp. Phys. \textbf{2014}, 073J01  (2014).

\bibitem{Kellogg}
P. J. Kellogg,
\textit{Van Allen Radiation of Solar Origin},
Nature {\bf 183}, 1295-1297 (1959).

\bibitem{Brice}
N. Brice and T. R. McDonough, 
\textit{Jupiter's radiation belts}, 
Icarus {\bf 18}, 206-219 (1973).

\bibitem{Coroniti}
F. V. Coroniti,
\textit{Energetic electrons in Jupiter's magnetosphere}, 
Atrophys J. Suppl. Ser. {\bf 27},261 (1974).

\bibitem{Nishida}
A. Nishida, 
\textit{Outward diffusion of energetic particles from the Jovian radiation belt},
J. Geophys. Res. {\bf 81} 1771-1773 (1976).

\bibitem{Carbary}
J. F. Carbary, T. W. Hill and A. J. Dessler, 
\textit{Planetary spin period acceleration of particles in the Jovian magnetosphere},
J. Geophys. Res. {\bf 81} 5189-5195 (1976).








\bibitem{Chen}
Y. Chen, D. R. Geoffrey and H. W. F. Reiner,
\textit{The energization of relativistic electrons in the outer Van Allen radiation belt},
Nature Phys. {\bf 3}, 614-617 (2007).

\bibitem{Horne}
R. B. Horne, R. M. Thorne, S. A. Glauert, J. D. Menietti, Y. Y. Shprits and D. A. Gurnett,
\textit{Gyro-resonant electron acceleration at Jupiter},
Nature Phys. {\bf 4}, 301-304 (2008).

\bibitem{whistler1}
S. J. Bolton, R. M. Thorne, D. A. Gurnett, W. S. Kurth and D. J. Williams, 
\textit{Enhanced whistler-mode emissions: Signatures of interchange motion in the Io torus}, 
Geophys. Res. Lett. {\bf 24}, 2123-2126 (1997). 

\bibitem{whistler2}
R. M. Thorne, T. P. Armstrong, S. Stone, D. J. Williams, R. W. McEntire, S. J. Bolton, D. A. Gurnett and M. G. Kivelson,
\textit{Galileo evidence for rapid interchange transport in the Io torus}, 
Geophys. Res. Lett. {\bf 24}, 2131-2134 (1997).

\bibitem{whistler3}
F. Xiao, R. M. Thorne, D. A. Gurnett and D. J. Williams, 
\textit{Whistler-mode excitation and electron scattering during an interchange event near Io}, 
Geophys. Res. Lett. {\bf 30}, 1749 (2003).

\bibitem{Olsen1981}
R. C. Olsen, 
\textit{Equatorially Trapped Plasma Populations},
J. Geophys Res. {\bf 86}, 235-245, (1981).

\bibitem{Olsen1987}
R. C. Olsen, S. D. Shawhan, D. L. Gallagher, J. L. Green, C. R. Chappell and R. R. Anderson,
\textit{Plasma Observations at the Earths Magnetic Equator},
J. Geophys Res. {\bf 92}, 2385-2407, (1987).

\bibitem{Persoon}
A. M. Persoon, D. A. Gurnett, O. Santolik, W. S. Kurth, J. B. Faden, J. B. Groene, G. R. Lewis, A. J. Coates, R. J. Wilson, R. L. Tokar, J.-E. Wahlund and M. Moncuquet,
\textit{A diffusive equilibrium model for the plasma density in Saturn's magnetosphere},
J. Geophys. Res. {\bf 114} A0421 (2009).

\bibitem{Yoshida2006}
Z. Yoshida, Y. Ogawa, J. Morikawa, S. Watanabe, Y. Yano, S. Mizumaki, T. Tosaka, Y. Ohtani, A. Hayakawa and M. Shibui,
\textit{First Plasma in the RT-1 Device},
J. Plasma Fusion Res. {\bf 1}, 008 (2006).

\bibitem{Yoshida2013}
Z. Yoshida, H. Saitoh, Y. Yano, H. Mikami, N. Kasaoka, W. Sakamoto, J. Morikawa, M. Furukawa and S. M. Mahajan,
\textit{Self-organized confinement by magnetic dipole: recent results from RT-1 and theoretical modeling}, 
Plasma Phys. Control. Fusion {\bf 55}, 014018 (2013).

\bibitem{Saitoh2014}
H. Saitoh, Y. Yano, Z. Yoshida, M. Nishiura, J. Morikawa, Y. Kawazura, T. Nogami, and M. Yamasaki,
\textit{Observation of a new high-beta and high-density state of a magnetospheric plasma in RT-1}, 
Phys. Plasmas {\bf 21}, 082511 (2014).

\bibitem{Boxer}
A. C. Boxer, R. Bergmann, J. L. Ellsworth, D. T. Garnier, J. Kesner, M. E. Mauel and P. Woskov,
\textit{Turbulent inward pinch of plasma confined by a levitated dipole magnet}, 
Nature Phys. {\bf 6}, 207 (2010).

\bibitem{Yoshida2010}
Z. Yoshida, H. Saitoh, J. Morikawa, Y. Yano, S. Watanabe, and Y. Ogawa,
\textit{Magnetospheric Vortex Formation: Self-Organized Confinement of Charged Particles},
Phys. Rev. Lett. {\bf 104}, 235004 (2010).

\bibitem{Saitoh2011}
H. Saitoh, Z. Yoshida, J. Morikawa, Y. Yano, T. Mizushima, Y. Ogawa, M. Furukawa, Y. Kawai, K. Harima, Y. Kawazura, Y. Kaneko, K. Tadachi, S. Emoto, M. Kobayashi, T. Sugiura and G. Vogel
\textit{High--$\beta$ plasma formation and observation of peaked density profile in RT-1}, 
Nucl. Fusion {\bf 51}, 063034 (2011).

\bibitem{Saitoh2015}
H. Saitoh, Y. Yano, Z. Yoshida, M. Nishiura, J. Morikawa, Y. Kawazura, T. Nogami, and M. Yamasaki,
\textit{Measurement of a density profile of a hot-electron plasma in RT-1 with three-chord interferometry}, 
Phys. Plasmas {\bf 22}, 024503 (2015).

\bibitem{Huba}
J. D. Huba, 
\textit{NRL Plasma Formulary}, 
(Naval Research Laboratory, Washington DC, 1994)

\bibitem{Schweer}
B. Schweer, G. Mank, A. Pospieszczyk, B. Brosda, and B. Pohlmeyer,
\textit{Electron temperature and electron density profiles measured with a thermal He-beam in the plasma boundary of TEXTOR},
J. Nucl. Mater. {\bf 196-–198}, 174 (1992).


\end{thebibliography}
\end{document}